\documentclass[12pt]{article}

\usepackage{epsf}

\newcommand{\rf}[1]{(\ref{#1})}
\newcommand{\bea}{\begin{eqnarray}}
\newcommand{\eea}{\end{eqnarray}}

\newcommand{\e}{{\rm e}}
\renewcommand{\d}{{\rm d}}

\renewcommand{\l}{\lambda}
\renewcommand{\L}{\Lambda}

\renewcommand{\a}{\alpha}
\newcommand{\n}{\nu}
\newcommand{\m}{\mu}

\newcommand{\del}{\delta}
\newcommand{\Del}{\Delta}

\newcommand{\ra}{\right\rangle}
\newcommand{\la}{\left\langle}
\newcommand{\prt}{\partial}
\newcommand{\vev}[1]{\left\langle {#1} \right\rangle}
\newcommand{\cD}{{\cal D}}

\newcommand{\tdqg}{two-dimensional quantum gravity}
\newcommand{\intm}{\int\!\!}
\def\void{}
\def\labelmark{}

\newenvironment{formula}[1]{\def\labelname{#1}
\ifx\void\labelname\def\junk{\begin{displaymath}}
\else\def\junk{\begin{equation}\label{\labelname}}\fi\junk}%
{\ifx\void\labelname\def\junk{\end{displaymath}}
\else\def\junk{\end{equation}}\fi\junk\labelmark\def\labelname{}}

{\ifx\void\labelname\def\junk{\end{array}\end{displaymath}}
\else\def\junk{\end{array}\right.\end{equation}}
\fi\junk\labelmark\def\labelname{}\def\junk{}
}

\newcommand{\beq}{\begin{formula}}
\newcommand{\eeq}{\end{formula}}
\newcommand{\beqv}{\begin{formula}{}}

\begin{document}
\topmargin 0pt
\oddsidemargin 5mm
\headheight 0pt
\headsep 0pt
\topskip 9mm

\begin{flushright}
NBI-HE-98--12\\
TIT/HEP--352\\
MPS-RR-1998-10\\
August  1998\\
\end{flushright}

\begin{center}
\vspace{24pt}
{ \large \bf Quantum Geometry and Diffusion}

\vspace{24pt}

{\sl J. Ambj\o rn}$\,^{a,}$\footnote{email ambjorn@nbi.dk} ,
{\sl K. N. Anagnostopoulos}$\,^{a,}$\footnote{email konstant@nbi.dk}, \\
{\sl T. Ichihara}$\,^{b,}$\footnote{email tomo@th.phys.titech.ac.jp}, 
{\sl L. Jensen}$\,^{a,}$\footnote{email ljensen@nbi.dk} and 
{\sl Y. Watabiki}$\,^{b,}$\footnote{email watabiki@th.phys.titech.ac.jp}

\vspace{24pt}

$^a$~The Niels Bohr Institute\\
Blegdamsvej 17, \\
DK-2100 Copenhagen \O , Denmark, 

\vspace{24pt}
$^b$~Department of Physics, \\
Tokyo Institute of Technology,\\ 
{\O}-okayama, Meguro, Tokyo 152, Japan

\end{center}
\vspace{24pt}

\vfill

\begin{center}
{\bf Abstract}
\end{center}

\vspace{12pt}

\noindent
We study the diffusion equation in two-dimensional quantum gravity,
and show that the spectral dimension is two despite the fact that the
intrinsic Hausdorff dimension of the ensemble of two-dimensional
geometries is very different from two. We determine the scaling
properties of the quantum gravity averaged diffusion kernel.

\vfill

\newpage

\section{Introduction}

Two-dimensional quantum gravity has been an interesting laboratory for
the study of fluctuating geometry. Many aspects have been understood
by means of Liouville field theory, matrix models and the transfer
matrix formulation of the theory. In particular, the transfer matrix
formulation \cite{transfer} has been useful for the analysis of what
we will call ``quantum geometry'', i.e.\ aspects of geometry which
have no classical analogy. Surprisingly, such a situation appears
already in pure two-dimensional quantum gravity. The partition
function for pure \tdqg\ where the volume of space-time is fixed to
$V$ is
\beq{1.1}
Z_V = \int\! \cD [g_{ab}] \;\del (\int\!\! \sqrt{g} -V),
\eeq
where $[g_{ab}]$ denote equivalence classes of metrics under
reparametrization.  With this definition the partition function for a
fixed cosmological constant can be written as the Laplace
transformation of $Z_V$:
\beq{1.2}
Z_\L = \int\!\! dV \; \e^{-\L V} Z_V.
\eeq
From \rf{1.1} it follows that each {\it geometry} $[g_{ab}]$ is
assigned the same weight (one), i.e.\ there is no classical minimum
around which it is natural to expand. This is why certain geometric
aspects related to $Z_V$ (or $Z_\L$) will be truly non-classical.  A
close analogy is found for the free relativistic particle. Let
$[P(x,y)]$ denote the equivalence class of paths from $x\in R^D$ to
$y\in R^D$, up to reparametrization invariance, and $L([P])$ the
length of the path in $R^D$. The propagator of the free particle has a
path integral representation closely analogous to \rf{1.1}--\rf{1.2}
\beq{1.3}
G_L (x,y) = \int\!\! \cD [P(x,y)] \;\del (L([P]) - L),~~~~~
G_M (x,y) = \int\!\! dL \; \e^{-ML} G_L(x,y).
\eeq
It is seen that each world-line contributes with weight one in the
path-integral representation of $G_L$, precisely as each geometry did
in the path-integral representation of $Z_V$. It is well-known that a
``typical'' path $[P(x,y)]$ has an (extrinsic) fractal dimension $D_H
= 2$. For instance, let us consider the ensemble of all equivalence
classes of paths of length $L$. The corresponding partition function
is
$$
G_L = \intm dx \; G_L(x,y),
$$
and we can calculate
\beq{1.4}
\la |x-y|\ra_L \equiv \frac{1}{G_L} \intm dx \, \cD[P(x,y)]\;
\del(L([P(x,y)])-L)\;  |x-y| \sim L^{1/2}.
\eeq
This is one of the simplest, but also most important quantum
phenomena: as long as we address distances less than the inverse
(renormalized) mass, it makes no sense to talk about any ordinary,
one-dimensional path of the particle.  Only for distances much larger
than the renormalized mass, one can talk about an approximate
classical path.

In the case of pure two-dimensional quantum gravity we have a somewhat
similar situation, only will geometries and fractal dimensions refer
entirely to intrinsic properties, with no reference to any embedding
space. Let $S_V(R)$ denote the average volume of a spherical shell of
geodesic radius $R$ in the ensemble of geometries with volume $V$. It
can be shown that \cite{aw,ajw}
\beq{1.5}
S_V(R) \sim R^3 (1 + O(R^7)).
\eeq
For any compact manifold of dimension $d$ and a given, smooth geometry
$[g_{ab}]$ we have that
\beq{1.5a}
S_V(R) \sim  R^{d-1}~~~~~{\rm for}~~~R \to 0.
\eeq
For an ensemble of geometries we call the power $d_h$ which appears
instead of $d$ for the {\it intrinsic fractal dimension} or the {\it
intrinsic Hausdorff dimension}. From \rf{1.5} it follows that $d_h =4$
for pure gravity, rather than $d_h =2$ as one would naively have
expected. Also, a calculation analogous to the one leading to \rf{1.4}
gives
\beq{1.6}
\la R \ra_V \sim V^{1/4},
\eeq 
which expresses that the average distance between two points in the 
ensemble of geometries only grows as $V^{1/4}$ and not as $V^{1/2}$.

The precise definition of $S_V(R)$ in pure quantum gravity is as 
follows: 
\beq{1.7}
S_V(R) = \frac{1}{Z_V} \; \frac{1}{V} 
\intm \cD [g_{ab}]\; \del(\intm \sqrt{g}\! -\!V) \int\!\! \intm
\sqrt{g(\xi_1)}\sqrt{g(\xi_2)}\; \del(D_g(\xi_1,\xi_2)-R),
\eeq
where $D_g(\xi_1,\xi_2)$ denotes the geodesic distance between the 
points labelled by $\xi_1$ and $\xi_2$. From the explicit calculation 
of $S_V(R)$ in pure gravity we know that \cite{aw,ajw} 
\beq{1.8}
S_V(R) = R^3 f\Big(\frac{R}{V^{1/4}}\Big),
\eeq
where $f(0) > 0$ and $f(x)$ falls off like $\e^{-x^{4/3}}$ for large
$x$.  Note that \rf{1.6} follows from \rf{1.8}.

From \rf{1.7} it follows that $S_V(R)$ can be viewed as a kind of
reparametrization invariant two-point function between points
separated by a geodesic distance $R$. The definition can be
generalized to include matter fields. For a given metric $g_{ab}$ the
reparametrization invariant partition function for matter will be
denoted $Z_{\rm m}[g_{ab}]$, and it will appear as a weight in \rf{1.1} and
\rf{1.3}. In this case it has not been possible to calculate $d_h$ by
the same constructive arguments which led to $d_h=4$ for pure
gravity. However, there exist arguments \cite{watabiki}, to be
reviewed in the next section, based on the diffusion equation in
quantum Liouville theory, which strongly suggest that $d_h(c)$ is a
non-trivial function of $c$ given by:
\beq{1.9}
d_h(c) = 2 \frac{\sqrt{25-c}+\sqrt{49-c}}{\sqrt{25-c}+\sqrt{1-c}}.
\eeq 
This formula agrees with the constructive approach for $c=0$ and 
for $c \to -\infty$ $d_h(c) \to 2$ as one would naively expect.
As we increase $c$ space-time becomes more fractal until the analytic 
formula breaks down for $c > 1$. For $c= -2$ we have a very precise 
verification of \rf{1.9} by numerical simulations \cite{many}. However,
for $0< c\leq 1$ the agreement with numerical simulations is not 
so good \cite{syracuse,ajw,new}. We will return to this question below.

While the definition of fractal dimension based on $S_V(R)$ is in
many ways natural, it is not the only one available. An alternative
definition is based on diffusion and the dimension defined
in this way is called the {\it spectral} dimension. The definition
has the advantage that it  makes sense when defined on
``fractal structures'' and we have just argued that a ``generic''
geometry in \tdqg\ in a certain sense {\it is} fractal. For a fixed 
(smooth) metric $g_{ab}$ the diffusion equation has the form:
\beq{1.10}
\frac{\prt}{\prt T} \, K_g(\xi,\xi_0;T) = \Del_g K_g (\xi,\xi_0;T),
\eeq
where $T$ is a fictitious diffusion time, $\Del_g$ is the Laplace
operator corresponding to the metric $g_{ab}$ and $K_g(\xi,\xi_0;T)$
denotes the probability density of diffusion from $\xi$ to $\xi_0$ in
diffusion time $T$. If we consider diffusion with the initial condition
\beq{1.11}
K_g(\xi,\xi_0;T=0) = \frac{1}{\sqrt{g(\xi)}}\, \del(\xi-\xi_0)
\eeq
it is well-known that $K_g$ has the following asymptotic expansion
for small $T$:
\beq{1.12}
K_g(\xi,\xi_0;T) \sim \frac{\e^{-D_g^2(\xi,\xi_0)/4T}}{T^{d/2}}
\sum_{r=0}^\infty a_r(\xi,\xi_0)\, T^r.
\eeq
In particular the average {\it return probability}
\beq{1.13}
RP_{g}(T) \equiv \frac{1}{V} \intm \sqrt{g(\xi)} \; K_g(\xi,\xi;T) \sim
\frac{1}{T^{d/2}} \sum_{r=0}^\infty A_r T^r,
\eeq
where 
$$
A_r = \frac{1}{V}\intm \sqrt{g(\xi)} \;a_r(\xi,\xi).
$$
The power $T^{d/2}$ reflects the dimension of the manifold, the
heuristic explanation being that small $T$ corresponds to small
distances and for any given smooth metric short distances mean flat
space-time.  However, the definition is more general, and can be
applied for diffusion in fractal structures, with the Laplacian
$\Delta_g$ appropriately defined, as is well-known from the theory of
percolation.  From \rf{1.12} we have (for a smooth metric $g_{ab}$) the
classical result
\beq{1.12a}
\frac{1}{V}\intm \intm \sqrt{g(\xi)}\sqrt{g(\xi_0) }
\; (D_g(\xi,\xi_0))^2 K_g(\xi,\xi_0;T) \sim T +O(T^2),
\eeq
irrespectively of $d$.

Since the probability $K_g$ is invariant under reparametrizations it
makes sense to define the quantum average of $K_g$ over all metrics:
\bea
K_V(R;T) &= &\frac{1}{Z_V}\; \frac{1}{S_V(R) V} \intm\cD [g_{ab}] \;
\del(\intm \sqrt{g}-V)\; Z_{\rm m}[g_{ab}] \nonumber \\ 
&& \times \intm\intm \sqrt{g(\xi)}\sqrt{g(\xi_0)} \; \del(D_g(\xi,\xi_0)\!
-\! R)\; K_g(\xi,\xi_0;T).\label{1.14}
\eea
By definition we have
\beq{1.15}
\int_0^\infty\! dR \, S_V(R)\; K_V(R;T) = 1,
\eeq
and furthermore, the quantum gravity average of $RP_g(T)$ is
\beq{1.16}
RP_V(T) = 
\frac{1}{Z_V} \intm \cD [g_{ab}] \;\del(\intm \sqrt{g}-V) \, 
Z_{\rm m}[g_{ab}] \, RP_g(T) 
= K_V(0;T).
\eeq

It natural to assume that $K_V(R;T)$ and $RP_V(T)$ have asymptotic
expansions somewhat like \rf{1.12} and \rf{1.13}. However, the powers
of $T$ which enter might be different from the canonical ones obtained
for a fixed, smooth geometry. This situation is well-known from the
study of diffusion on fixed fractal structures (see \cite{review} for
a review).  One operates with two different exponents. A dynamical
exponent (or dimension) $\delta_w$ related to diffusion (or random walk) on
the fractal structures and a structural dimension, which we here
identify with the intrinsic Hausdorff dimension\footnote{In the study
of diffusion on fixed fractal structure one usually imagines the
fractal structure embedded in $R^D$. Thus one has an extrinsic fractal
dimension $D_H$ and intrinsic fractal dimension $d_h$, the last one
defined with respect to the ``geodesic distance'' of the fractal,
which is defined from the shortest path between to points on the
fractal. One usually has $d_h = \tilde{\n} D_H$ for some positive
constant $\tilde{\n}$. In the same way one has a relation similar to
\rf{1.16a}, only with the distance $R_E(T)$ measured in $R^D$, rather
than intrinsically on the fractal:
$$
\la R_E^2(T) \ra_V \sim T^{2/\Delta_W}.
$$
The exponent $\delta_w = \tilde{\n} \Delta_W$.} 
$d_h$. The exponent $\delta_w$ is defined by the mean-square displacement
after time $T$: 
\beq{1.16a}
\la R^2(T) \ra_V \sim T^{2/\delta_w},
\eeq
assuming that $R(T) \ll V^{1/d_h}$. This means that the volume 
covered by diffusion after time $T$ will be 
$V(T) \sim \la R(T) \ra_V^{d_h}$, and the probability that the random 
walk will return to the origin should behave as:
\beq{1.16c}
RP_V(T) \sim \frac{1}{T^{d_h/\delta_w}}(1+o(T)) \equiv \frac{1}{T^{d_s/2}}
(1+o(T)).
\eeq
Thus we have, by definition, 
\beq{1.16d}
d_s = \frac{2d_h}{\delta_w}.
\eeq
Our task in \tdqg\ is to determine two of the three quantities $d_h$, 
$d_s$ and $\delta_w$.

In the theory of diffusion on fractal structures it is usually assumed 
(and well established numerically) that in the limit $V\!\to\!\infty$ 
$K_\infty (R;T)$ has the following functional form
\beq{1.16e}
K_\infty (R;T) = \frac{1}{T^{d_s/2}} 
 \tilde{F}_\infty\Big( \frac{R}{T^{1/\delta_w}} 
\Big),
\eeq
where $\tilde{F}_\infty(x)$ falls off approximately as $\e^{-x^u}$. 
Various values of $u$ has been considered, ranging from $u=1$ to $u = 
\delta_w/(\delta_w-1)$. The functional form \rf{1.16e} of course reproduces
\rf{1.16a} since 
\beq{1.16f}
\la R^n(T) \ra_\infty \sim \intm \d R\,  R^{d_h-1} \; R^n \; 
K_\infty (R;T) \sim ~T^{n/\delta_w}.
\eeq
In the case of \tdqg\ we want to consider a fixed volume $V$ and
average over all shapes. The original heat kernel expansion for a
fixed geometry contains reference to powers of the curvature, but
since we integrate over all geometries one expects that only reference
to $V$ will survive. We thus conjecture that
\beq{1.16g}
V \, K_V(R;T) = \frac{V}{T^{d_s/2}} \tilde{F}\Big( \frac{R}{T^{1/\delta_w}}, 
\frac{T}{V^{2/d_s}}\Big)= 
\frac{V}{T^{d_s/2}} {F}\Big( \frac{R}{V^{1/d_h}}, \frac{T}{V^{2/d_s}}
\Big),
\eeq
where we have used \rf{1.16d} to write
\beq{extra}
\tilde{F}
 \Bigg( \frac{R}{T^{\frac{1}{\delta_w}}}, \frac{T}{V^{\frac{2}{d_s}}}\Bigg)=
\tilde{F}
 \Bigg( \frac{R}{V^{\frac{1}{d_h}}}
  \Bigg[\frac{T}{V^{\frac{2}{d_s}}}\Bigg]^{\frac{-1}{\delta_w}}, 
  \frac{T}{V^{\frac{2}{d_s}}}\Bigg) \equiv 
       F 
 \Bigg(\frac{R}{V^{\frac{1}{d_h}}},\frac{T}{V^{\frac{2}{d_s}}}\Bigg).
\eeq
We expect the following boundary conditions on $F$ and $\tilde{F}$: 
\bea
{F}(x,y) &\sim& y^{d_s/2} ~~~~~{\rm for}~~y\to \infty,
\label{1.16h}\\
\tilde{F}(x,y) &\to & 0~~~~~~~~~~{\rm for}~~y \to \infty,~~~~x >0, 
\label{1.16i}\\
\tilde{F}(x,y) &\to & \tilde{F}_\infty(x)~~~{\rm for}~~ y\to 0.
\label{1.16j}
\eea
\rf{1.16h} results from the fact that $K_V (R;T) \to 1/V$ for $T \to
\infty$. We obtain \rf{1.16j} from \rf{1.16e} and \rf{1.16i} because
of \rf{1.11}. The above conditions are verified in our numerical
simulations. 

Finally the return probability for a finite volume will be given by 
\beq{1.16k}
RP_V(T) = \frac{1}{T^{d_s/2}}F\Big(0,\frac{T}{V^{2/d_s}}\Big).
\eeq
These are the scaling ans\"{a}tze we will use in the following.
Note that they imply the following:
\bea
\la R^n(T) \ra_V &\sim& 
 T^{n/\delta_w}~~~~~{\rm for}~~~T \to 0,\label{1.16l}\\
\la R^n(T)\ra_V &\sim& 
 V^{n/d_h} ~~~~~{\rm for}~~~T \to \infty. 
\label{1.16m}
\eea 

For any fixed, smooth geometry $[g_{ab}]$ $d_h = d_s =d$, where $d$
denotes the dimension of the underlying manifold (i.e.\ equal 2 in
\tdqg). After taking the functional average over geometries we know
that $d_h$ changes, as already discussed.  However, {\it we will
provide evidence that $d_s$ is unchanged and equal to two for all
values $c \le 1$ of the central charge $c$ of the matter fields coupled
to quantum gravity}.  In this context it is worth to recall that there
exists a recent analytical argument in favour of this scenario
\cite{d_sana}: for Gaussian fields $X^\m(\xi_1,\xi_2)$, $\m =
1,\ldots,D$, coupled to two-dimensional gravity it is possible to
derive the following relationship between the {\it extrinsic}
Hausdorff dimension $D_H$ of the surface $X^\m(\xi_1,\xi_2)$ embedded
in $R^D$ and the spectral dimension $d_s$,
\beq{newx1}
d_s = \frac{2D_H}{D_H+2}.
\eeq

One assumption going into this derivation is the scaling ansatz
\rf{1.16g} for $R=0$. Next, assuming that we can perform an analytic
continuation of $D$ from positive integers to $D \in \, ]-\infty,1[$ one
can appeal to Liouville theory and argue that $D_H = \infty$ for these
values of $D$. Thus $d_s= 2$ for this particular model. The numerical
experiments reported in this article will provide evidence that the
result $d_s=2$ is of larger generality than what was proven in
\cite{d_sana}, and also provide support for the scaling
\rf{1.16g}. Note that for branched polymers (where $D_H=4$) \rf{newx1}
results in  $d_s=4/3$ in agreement with \cite{jw}.

The rest of this article is organized as follows: In section
\ref{liouville} we review shortly the diffusion in quantum Liouville
theory and the derivation of \rf{1.9}, while section \ref{numeric}
presents the numerical evidence for scaling.  Finally section
\ref{discussion} contains a discussion of the results obtained.

\section{Diffusion in Liouville theory}\label{liouville}

Let $\Phi_n[g_{ab}]$ be a general spin-less operator which depends only on
the metric $g_{ab}$, is reparametrization invariant, and satisfies
$\Phi_n[\l g_{ab}] = \l^{-n}\Phi_n[g_{ab}]$ at the classical level.  The
expectation value of this operator in the context of \tdqg\ coupled to
a conformal field theory with central charge $c$ is defined by
\beq{2.1}
\la \Phi_n [g_{ab}]\ra_V = 
\frac{1}{Z_V} \intm \cD [g_{ab}] \; \del(\intm \sqrt{g}-V) \, 
Z_{\rm m}[g_{ab}]\; \Phi_n[g_{ab}].
\eeq
It follows  from Liouville theory (see \cite{watabiki}
for details) that we have the following scaling:
\beq{2.2}
\la \Phi_n[g_{ab}]\ra_{\l V} = \l^{\a_{-n}/\a_1} \la \Phi_n[g_{ab}]\ra_V,
~~~~~~
\a_n= \frac{2n\sqrt{25-c}}{\sqrt{25-c}+\sqrt{25-c -24n}}.
\eeq
Consider now the diffusion kernel $K_g (\xi,\xi_0;T)$  discussed
in the introduction. The formal solution is given as
\beq{2.3}
K_g (\xi,\xi_0;T) = \e^{T \Del_g(\xi)} K_g(\xi,\xi_0;0).
\eeq
We get the return probability by setting $\xi=\xi_0$ (after acting
with $\e^{T \Del_g (\xi)}$) and taking the average over all $\xi_0$.
If we expand in $T$ we obtain:
\beq{2.4}
K_g (\xi,\xi;T) = \left[( 1 + T \Del_g + \cdots) \frac{1}{\sqrt{g(\xi)}}
\del(\xi-\xi_0)\right]_{\xi_0 = \xi}.
\eeq
Let us assume the existence of a $T'$ such that
\beq{2.5}
\l V RP_{\l V} (T') = V RP_V (T).
\eeq
From the assumed scaling ansatz \rf{1.16k} it follows that
\beq{2.5a}
T' = \l^{2/d_s} T = \l^{\delta_w/d_h} T.
\eeq
Since the scaling properties of the operator $\Del_g$ will change
when dressed by \tdqg, it is clear that one cannot maintain the 
combination $T \Del_g$ in \rf{2.3} and \rf{2.4} with $T$ having it's
canonical dimension after averaging over 
all geometries. A better guess is obtained as follows: 
the average of the square of the geodesic distance travelled
by diffusion at time $T$ for a fixed geometry,
as defined in \rf{1.12a}, is again a reparametrization invariant
object, and it makes sense to define the average in the ensemble
of two-dimensional geometries weighted by $Z_{\rm m}[g_{ab}]$. 
Naively, one would expect
\beq{2.7}
\la \frac{1}{V} \intm \intm \sqrt{g(\xi)}\sqrt{g(\xi_0)}
\; D_g^2(\xi,\xi_0) \;K_g(\xi,\xi_0;T) \ra_V  \sim T + O(T^2),
\eeq 
the first term 
proportional to $T$ coming from $T \Del_g$ if we use the expansion
\rf{2.4}, and since we {\it might} expect a more general expression
\rf{1.16l} after averaging over geometries, it is natural to 
assume that one should consider the combination $T^{2/\delta_w}\Del_g$.
Then the first term in the expansion \rf{2.4} would reproduce
the behaviour \rf{1.16l}, while \rf{2.4} and \rf{2.2} and 
the scaling properties of $\Del_g$ would allow us to conclude that 
\beq{2.5b}
{\rm dim} \left[ T^{2/\delta_w}\right] = 
{\rm dim} \left[ V^{-\a_{-1}/{\a_1}} \right].
\eeq
From \rf{1.16l} and \rf{1.16m} we arrive at the formula \rf{1.9}:
\beq{2.9}
d_h = -\frac{2\a_1}{\a_{-1}} = 
2 \frac{\sqrt{25-c}+\sqrt{49-c}}{\sqrt{25-c}+\sqrt{1-c}}.
\eeq
However, note that this kind of argument does not allow us to determine
the dimension of $T$, i.e.\ $d_s$ or $\delta_w$. This will be the purpose 
of the rest of the article.

\section{Numerical methods and results}\label{numeric}

\begin{figure}[htb]
\centerline{\epsfxsize=4.0in \epsfysize=2.67in 
 \epsfbox{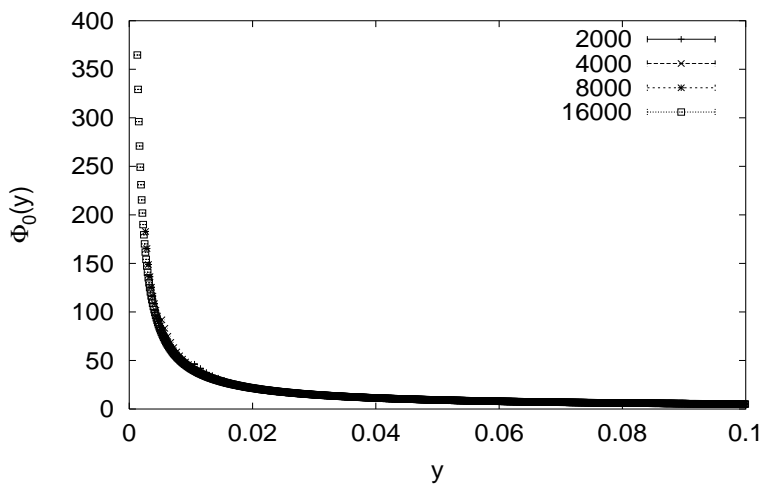}}
\caption{Finite size scaling of the return probability for the Ising
model.}
\label{f:1}
\end{figure}

\begin{figure}[htb]
\centerline{\epsfxsize=4.0in \epsfysize=2.67in 
 \epsfbox{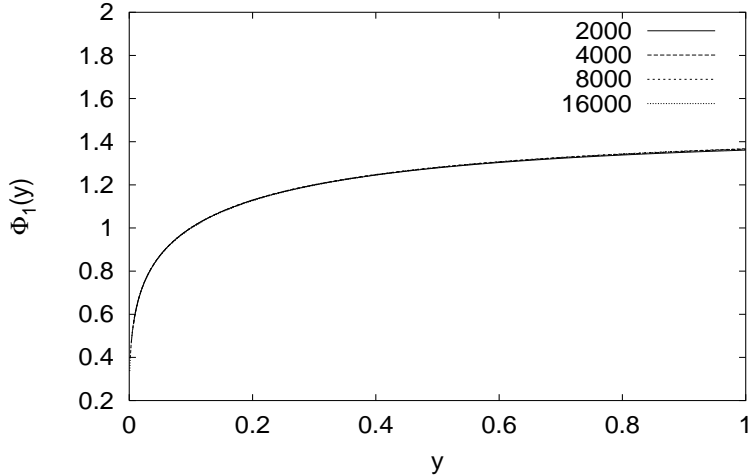}}
\caption{Finite size scaling of the average distance travelled by the
random walker for the Ising model.}
\label{f:2}
\end{figure}

\begin{table}[ht]
\begin{center}
\begin{tabular}{|c|l|l|l|l|l|l|}
\hline
\multicolumn{1}{|c} {     }&
\multicolumn{2}{|c} {$n=0$}&
\multicolumn{2}{|c} {$n=1$}&
\multicolumn{2}{|c|}{$n=2$}\\
\cline{2-7}
\multicolumn{1}{|c} {$c$}  &
\multicolumn{1}{|c} {$d_s$}&
\multicolumn{1}{|c} {$b$}  &
\multicolumn{1}{|c} {$d_h$}&
\multicolumn{1}{|c} {$a$}  &
\multicolumn{1}{|c} {$d_h$}&
\multicolumn{1}{|c|}{$a$}  \\
\hline
$-2$ &  2.00(3)&   2(5)& 3.58(13)&   0.6(3)& 3.59(12)& 0.6(3)\\
0    & 1.991(6)&   4(5)& 4.09(23)&   1.2(6)& 4.08(25)& 1.1(5)\\
1/2  & 1.989(5)&   4(4)& 4.08(32)&   0.9(5)& 4.09(28)& 0.9(5)\\
4/5  & 1.991(5)&   5(5)& 3.99(24)&   0.7(5)& 3.98(18)& 0.7(5)\\
\hline
\end{tabular}

\bigskip

\begin{tabular}{|c|l|l|l|l|}
\hline
\multicolumn{1}{|c} {     }&
\multicolumn{2}{|c} {$n=3$}&
\multicolumn{2}{|c|}{$n=4$}\\
\cline{2-5}
\multicolumn{1}{|c} {$c$}  &
\multicolumn{1}{|c} {$d_s$}&
\multicolumn{1}{|c} {$a$}  &
\multicolumn{1}{|c} {$d_h$}&
\multicolumn{1}{|c|}{$a$}  \\
\hline
$-2$ &  3.55(9)& 0.5(3)&  3.53(8)&   0.4(2)                 \\
0    & 4.10(20)& 1.2(4)& 4.10(15)&   1.2(5)                 \\
1/2  & 4.10(25)& 1.0(4)& 4.11(23)&   1.0(4)                 \\
4/5  & 3.98(16)& 0.7(4)& 3.98(14)&   0.7(4)                 \\
\hline
\end{tabular}
\end{center}
\caption{The values for the spectral dimension $d_s$ and the fractal
dimension $d_h$ obtained from calculating $\Phi_n(y)$ using finite
size scaling. The sizes of the lattices are $N=$ 2, 4, 8 and 16K
triangles.}
\label{t:1}
\end{table}

\begin{table}[ht]
\begin{center}
\begin{tabular}{|c|l|l|l|}
\hline
\multicolumn{1}{|c} {$N$}       &
\multicolumn{1}{|c} {$d_s$}     &
\multicolumn{1}{|c} {$b$}       &
\multicolumn{1}{|c|}{$d_s(b=0)$}\\
\hline
128000& 1.980(14) & 1.0(7)  &  1.9586(4)\\
64000 & 1.972(18) & 0.9(5)  &  1.9400(5)\\
32000 & 1.958(32) & 0.8(8)  &  1.925(1) \\
16000 & 1.954(18) & 0.9(3)  &  1.8949(7)\\
8000  & 1.938(34) & 0.8(4)  &  1.865(2) \\
4000  & 1.934(58) & 0.9(4)  &  1.826(3) \\
\hline
\end{tabular}
\end{center}
\caption{The values of the spectral dimension $d_s$ obtained from the
small time scaling of $\Phi_0(y)$ for the $c=-2$ model.}
\label{t:2}
\end{table}

\begin{figure}[htb]
\centerline{\epsfxsize=4.0in \epsfysize=2.67in \epsfbox{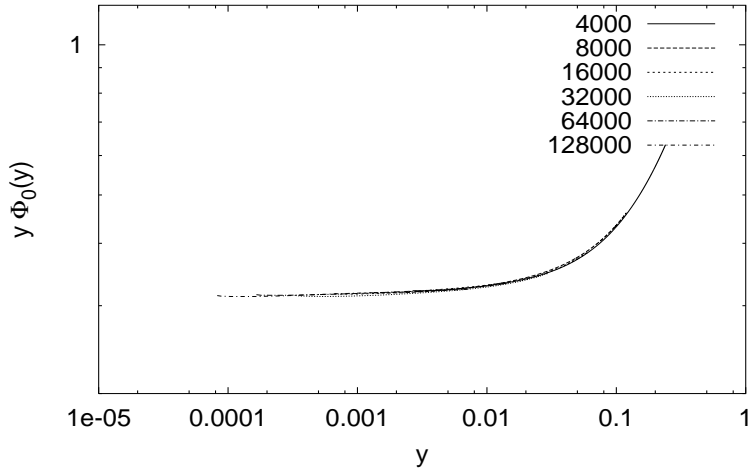}}
\caption{Small time behaviour of the return probability for the $c=-2$
model.}
\label{f:3}
\end{figure}

\begin{figure}[htb]
\centerline{\epsfxsize=4.8in \epsfysize=2.67in \epsfbox{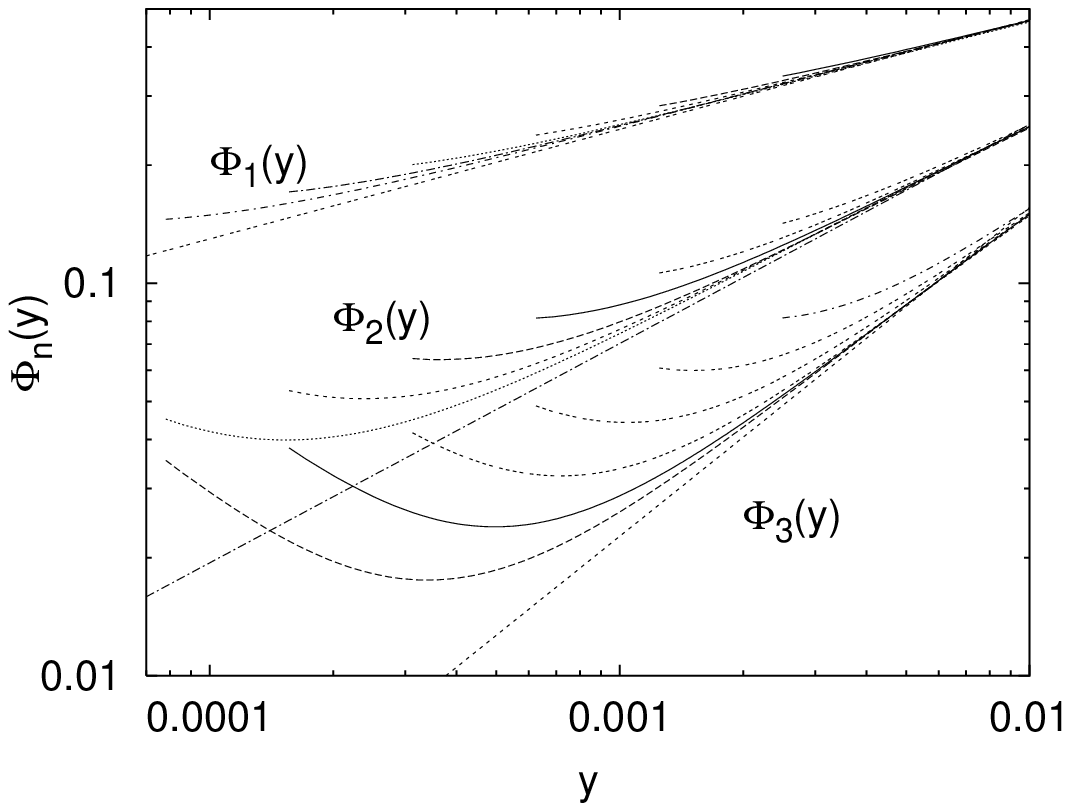}}
\caption{Small time behaviour of the moments $\Phi_n(y)$ for the
$c=-2$ model. $N=$ 4--128K from right to left for each
distribution. The straight lines are $\propto y^{nd_s/2d_h}$ for
$d_s=2$ and $d_h=3.58$. Slow convergence is observed as $N\to\infty$.}
\label{f:4}
\end{figure}

We used dynamical triangulations in order to simulate conformal matter
coupled to 2d quantum gravity on the lattice. In this approach, a
triangulation $\cal T$ corresponds to an equivalence class of metrics
$[g_{ab}]$ in \rf{1.1} and the volume $V$ of spacetime is given by the
number of triangles $N$ in $\cal T$. We used standard Monte Carlo
techniques for unitary matter with $c=0$ (pure gravity), $1/2$ (Ising
model) and $4/5$ (3--states Potts model) and an effective recursive
sampling technique for the (non--unitary) $c=-2$ model which
constructs independent configurations. Details of the methods and
models we used can be found in \cite{many,new}. A certain number of
configurations was generated and the diffusion field 
$K_N^{\cal T}(P,P_0;T)$, defined on vertices, 
was evolved using the discretized version of \rf{1.10}
\beq{3.1}
K_N^{\cal T}(P,P_0;T+1)=
 \sum_{j} \frac{1}{n(P_j)}K_N^{\cal T}(P_j,P_0;T), \qquad
K_N^{\cal T}(P,P_0;0)=\delta_{P,P_0}\, ,
\eeq
where $j$ runs over the neighbours of $P$ and $ n(P_j)$ denotes the
connectivity number of the vertex $P_j$. Then we obtain
\beq{3.2}
K_N(R;T)=\frac{1}{S_N(R)}\vev{\sum_P \delta_{d_{\cal T}(P,P_0),R}
K_N^{\cal T}(P,P_0;T)}_{\cal T}\, .
\eeq
$d_{\cal T}(P,P_0)$ is the geodesic distance (shortest link path)
between the points $P$ and $P_0$ and $S_N(R)$ is the number of vertices 
at geodesic distance $R$ from the point $P_0$. The return probability
$RP_N(T)$ and the moments $\vev{R^n(T)}_N$ can easily be calculated
\bea
RP_N(T) &=& \vev{K_N^{\cal T}(P_0,P_0;T)}_{\cal T} \, ,
    \label{3.3a}\\
\vev{R^n(T)}_N&=&
  \sum_{R=0}^\infty R^n S_N(R) K_N(R;T)
    \label{3.3b}\\
              &=&
  \vev{\sum_P d_{\cal T}(P,P_0)^n \,K_N^{\cal T}(P,P_0;T)}_{\cal T}
    \label{3.3c}\, .
\eea
It is more convenient to use \rf{3.3c} for calculating
$\vev{R^n(T)}_N$. Evolving the field turns out to be an expensive
procedure, so only one point $P_0$ was chosen per configuration. For
this reason, the sampling of configurations in the case of unitary
models was done sufficiently far apart so that they were essentially
independent from each other.

From the scaling relations \rf{1.16a}--\rf{1.16m}, one expects that
$RP_N(T)$ and  $\vev{R^n(T)}_N$  will be functions of the scaling
variables 
\bea
x &=& \frac{R+a}{N^{1/d_h}}\label{3.4a} \, , \\
y &=& \frac{T+b}{N^{2/d_s}}\label{3.4b}\, ,
\eea
where the ``shifts'' $a$ and $b$ are the lowest order finite size
corrections to scaling. The shift $a$ has been used with great success
in measuring correlation functions as functions of the geodesic
distance \cite{ajw,many,new} improving dramatically their scaling
properties and making it possible to probe the fractal structure even
for moderately small lattices. We expect such corrections to be
necessary in our case as well. The scaling relations to be tested in
the simulations are
\bea
RP_N(T) &=& \frac{1}{N} \Phi_0(y)\label{3.5a} \, , \\
\vev{R^n(T)}_N &=& N^{n/d_h} \Phi_n(y)\label{3.5b}\, .
\eea
As we will see, from \rf{3.5a} one obtains $d_s\approx 2$ for all
models.  Using this value in \rf{3.5b}, one obtains values for $d_h$
consistent with the ones measured using different observables
\cite{many,new} verifying this way eq.~\rf{1.16d}.

The functions $\Phi_n(y)$ for small $y$ are expected to behave as:
\bea
\Phi_0(y) &\sim& y^{-d_s/2}\label{3.6a} \, , \\
\Phi_n(y) &\sim& y^{n d_s/2 d_h}\, ,\qquad n>0\label{3.6b}\, .
\eea

The validity of \rf{3.5a} and \rf{3.5b} was tested by collapsing the
distributions for different values of $N$ ranging from 2--16K
triangles.  The method is described in full detail in
\cite{many,new}. Measurement time grows as $N^2$ and this puts a
severe limit on the maximum size of configurations possible to be
studied. Measurements were performed on approximately 50000
configurations (10000--14000 for the 16K lattice).  For the unitary
models a configuration was obtained every 100 sweeps.  One point $P_0$
was randomly chosen on each configuration. The best values for $d_s$
and $d_h$ are recorded in Table~\ref{t:1}. In the case of
$\vev{R^n(T)}_N$, $d_s$ was fixed to be equal to $2$, $b$ was set to 0
and the fractal dimension $d_h$ as well the shift $a$ were the free
parameters to be tuned.  The introduction of the shift $a$ is crucial
for these functions to collapse reasonably. In the case of $RP_N(T)$
the only parameters involved are the $T$-shift $b$ and the spectral
dimension $d_s$. It is worth mentioning that the collapse was done for
a wide range of $y$ (0.01--1) and that $(\chi^2/{\rm dof})_{\rm min}$
was considerably less than one in all cases (0.2-0.5 for approx. 45000
dof in each group). The errors quoted are for the range where
$\chi^2/{\rm dof}=1$. Figures~\ref{f:1} and \ref{f:2} show how well
the scaling relations hold in the case of the Ising model. Similar
figures can be obtained for the other models as well.

Measuring $d_s$ and $d_h$ using the small time behaviour \rf{3.6a},
\rf{3.6b} is more difficult. Larger lattices and more statistics are
necessary for a sensible measurement to be made. For a detailed study,
we confined ourselves to the $c=-2$ model where it is easy to generate
large configurations. Measurements were made on approximately 80000
configurations (41000 for the 128K lattice).  We evolved the diffusion
field up to $T=1000$ for lattices with $N=4K$--$128K$. We checked that
the results were consistent with the measurements we obtained from the
unitary models, although with much less accuracy.

In the case of $\Phi_0(y)$ the fits were performed by introducing the
shift $b$ and then by making a log-log plot for small $y$. The value
of $\chi^2/{\rm dof}$ was determined for a range of $b$ from which we
computed the best values of $d_s$ and $b$ and their errors quoted in
Table~\ref{t:1}. The small $T$ cutoff $T_{\rm min}=7$ was fixed for
all volumes such that it would be the smallest $T_{\rm min}$ giving
$\chi^2/{\rm dof}$ of order 1 for a reasonable range of $T$. The upper
limit was fixed $y_{\rm max}$ (i.e. $T_{\rm max}\propto N$). The fits
were reasonably stable with different choices of $T_{\rm min}$ and
$T_{\rm max}$.  The values of $d_s$ for $b=0$ for the same $T$-range
are also shown in Table~\ref{t:1} for comparison. We see that $b$
improves the value for $d_s$ quite a lot for the small lattices. In
Figure~\ref{f:3} we show graphically that \rf{3.4a} holds with $d_s=2$
with very good accuracy.

The above method is not so successful in the case of $\Phi_n(y)$ for
$n>0$. We observe large finite size effects entering in the
calculation, which grow with $n$, as can be seen in
Figure~\ref{f:4}. The straight lines correspond to the expected slopes
and we see a very slow convergence as $N\to\infty$.  The fits, even
for $n=0$, do not yield stable values for $d_h$ and the results depend
strongly on the range of $T$ chosen. One has to throw away several
small $T$ points in order to obtain reasonable values of
$\chi^2$. Finite size effects enter in eq.~\rf{1.16f} through the
assumption that $S_N(R)\approx R^{d_h-1}$ (which we know that for the
size of the surfaces studied is valid only for quite small values of
$R$) and from the assumption that $K_N(R;T)\approx \tilde{\Phi}_0(z)$
where $z\equiv R/T^{1/\delta_w}$ which holds only for $N\to\infty$.
\section{Discussion}\label{discussion}

The numerical results reported above are two-fold: a corroboration of
the conjecture that $d_s$ = 2 for conformal matter coupled to
two-dimensional quantum gravity, and a test of the scaling conjecture
\rf{1.16d} and \rf{1.16g}. The test of $d_s=2$ was two-fold.  For
$c=-2, 0, 1/2$ and $4/5$ a measurement of the return probability
allowed a test of the functional form of $V K_V(0;T)$ in the form
\rf{1.16g}, and in this way a determination of $d_s$. For a given
central charge $c$ it was done by ``collapsing'' the measurements for
various $V$ of the return probability as a function the single scaling
variable
\beq{final1}
y = \frac{T}{V^{2/d_s}}.
\eeq
This is possible with impressive accuracy for a wide range of $V$'s
and $T$'s if $d_s \approx 2$ as described above. The second,
independent, test was only carried out for $c=-2$ and concentrated on
the small $T$ dependence of $K_V(0;T)$. According to \rf{1.16g} a fit
to the power fall off should allow a determination of $d_s$. This
approach was used in the first systematic investigation of diffusion
in the context of two-dimensional quantum gravity \cite{ajw}. It does
not allow a determination of $d_s$ with the same precision as the
``collapse'' method, but has the advantage, from the point of view of
computer resources, that one only need to evolve the diffusion process
for a small time interval. Again the result is $d_s \approx 2$.

The final test of the scaling form \rf{1.16g} is performed under the
assumption that $d_s=2$. A measurement of the moments $\la R^n(T)
\ra_V$ allows a test of the $R$--dependent part of the scaling
hypothesis \rf{1.16g}. Again it is done by ``collapse'' of the
measured distributions of $\la R^n(T) \ra_V$ for various values of $T$
and $V$ and we find that their scaling is consistent with the
existence of a scaling variable
\beq{final2}
x = \frac{R}{V^{1/d_h}}
\eeq
over {\it all} scales on the surface. This is in agreement with
measurements on different correlation functions like the loop--length
distribution function \cite{transfer,many,new} (from which one obtains
e.g. $S_V(R)$).  It {\it is} possible to perform such a collapse for a
narrow range of $d_h$. In this way one obtains an independent
measurement of $d_h$, compared to the one obtained in
\cite{syracuse,ajw}.  The agreement with the $d_h$ obtained by a
direct measurement of the intrinsic Hausdorff dimension is perfect.
Alternatively, the consistency of the results can be seen as a
confirmation of \rf{1.16d}.

Summarizing, we have verified that with high accuracy  
$d_s=2$. Further, the scaling relation \rf{1.16g}
seems to be valid. Thus, we have a remarkable
situation: a generic geometry which appears in 
the path integral in two-dimensional quantum gravity,
is fractal with an intrinsic Hausdorff dimension $d_h$
(which is a function of the central charge $c$ of the 
matter coupled to gravity). On such a ensemble of geometries diffusion 
is ``anomalous'', i.e. 
\beq{final3}
\la R^2(T) \ra_V 
 \sim T^{2/\delta_w}(1 + \cdots)~~~~{\rm for}~~~T \ll V^{1/d_h},
\eeq
rather that $R^2(T) \sim T$ as in ordinary diffusion on a fixed 
smooth geometry. However, this anomalous diffusion is
counteracted by the fact that the geodesic distance
itself has an anomalous dimension, and if the only 
measure of diffusion was the return probability, 
such a fractal space-time geometry would appear 
indistinguishable from an ordinary smooth 
two-dimensional space-time geometry. 
  
The values of $d_h$ measured by diffusion agrees with the values
determined so far by direct geometric measurement
\cite{syracuse,ajw,many,new}.  In particular one observes perfect
agreement with \rf{1.9} for $c=-2$ and $c=0$. However, for $c=1/2$ and
$c=4/5$, i.e.\ in the case of unitary matter coupled to gravity, there
is not a very impressive agreement.  Thus we are still left with one
of the few remaining puzzles in two-dimensional quantum gravity: is
$d_h=4$ for the central charge $c \in [0,1]$, or does it follow the
prediction \rf{1.9} in this range of $c$ (as seems to be the case for
$c \leq 0$)?  The fact that several independent ways of measuring
$d_h$ agree, and fail to confirm \rf{1.9}, indicate that either
\rf{1.9} {\it is} not valid for $c > 0$, or there is a very general
reason for the failure of the numerical simulations.  One such reason
could be that the volumes $V$ considered so far are too small. Indeed,
there have been arguments in favour of large finite size effects for $c
>0$ \cite{hari}, but it is difficult for us to understand that one
then should be able to measure critical exponents of, say, the Ising
model coupled to quantum gravity with great accuracy, if
this model suffers severe finite size effects for the same volumes
when it comes to measurements of geometry. In particular, it is
difficult to understand such a discrepancy between finite size effects
on critical exponents and geometry when it is believed that it is the
fractal geometry which is responsible for the change in critical
exponents of the Ising model from the values in flat space to the KPZ
values \cite{abt}.

\section*{Acknowledgements}
J.A. acknowledges the support from MaPhySto which is financed by the 
Danish National Research
Foundation. Y.W. acknowledges the support 
and the hospitality of the Niels Bohr Institute.

\end{document}